\documentclass[a4paper,10pt]{article}
\usepackage{amsfonts}
\usepackage{amsmath}
\usepackage{amssymb}
\usepackage{slashed}
\usepackage[dvips]{graphicx}

\DeclareGraphicsExtensions{.bmp,.png,.pdf,.jpg}

\begin{document}

\title{Entanglement harvesting in double-layer graphene by vacuum
fluctuations in a microcavity}
\author{Juan Sebasti\'{a}n Ardenghi$^{1,2}$\thanks{%
email:\ jsardenghi@gmail.com, fax number:\ +54-291-4595142} \\
$^{1}$Departamento de F\'{\i}sica, Universidad Nacional del Sur, Avenida
Alem 1253,\\
B8000CPB, Bah\'{\i}a Blanca, Argentina \\
$^{2}$Instituto de F\'{\i}sica del Sur (IFISUR, UNS-CONICET), Avenida Alem
1253,\\
B8000CPB, Bah\'{\i}a Blanca, Argentina \\
}
\maketitle

\begin{abstract}
The aim of this work is to study the entanglement harvesting between two
graphene layers inside a planar microcavity. Applying time-dependent
perturbation theory it is shown that nonclassical correlations between
electrons in different layers are obtained through the exchange of virtual
photons. Considering different initial states of the electrons and the
vacuum state of the electromagnetic field, the negativity measure that
quantifies the entanglement, is computed through the photon propagator, for
time scales smaller than the light-crossing time of the double layer. The
results are compared with those obtained for hydrogenic probes and pointlike
Unruh-deWitt detectors, showing that for different initial states, entangled 
$X$ states and more general entangled reduced matrices are obtained, which
enlarge the classification of bipartite quantum states.
\end{abstract}

\section{Introduction}

The vacuum state of a free quantum field, contains correlations of different
observables in separate region of spacetime, even when those regions are
spacelike separated (\cite{valentini}, \cite{resnik}, \cite{pozak}). This
nonclassical behavior of the vacuum state of the field is a vital concept in
phenomena such as quantum collect calling \cite{jon}, the black hole
information loss problem (\cite{haw}, \cite{al}) and quantum energy
teleportation (\cite{hotta}, \cite{jsa1} and \cite{jsa2}). These
correlations are, in principle, physically accessible because they can be
obtained from the field vacuum via quantum particles detectors that couple
to it locally (\cite{reznik2}). This allows to observe an entanglement of
the particle detectors that are operated by observers, even if they remain
spacelike separated during their whole existence \cite{resnik}. The
phenomenon of extraction of nonclassical correlations from the quantum
vacuum has become known as entanglement harvesting, which was introduced in 
\cite{valentini}. Entanglement harvesting from scalar fields has been widely
studied \cite{pozak} and applied in entanglement farming \cite{mar1},
metrology \cite{salt} and in cosmology, where it has been shown that
entanglement harvesting is very sensitive to the geometry of the underlying
spacetime (\cite{stee} and \cite{mar2}) or even its topology \cite{mar3}. In
general, the detector-field interaction is modeled using the Unruh-DeWitt
model \cite{dewitt}, which consists of a linear coupling of a pointlike
two-level quantum system and a massless (or not), scalar quantum field,
where a spatial smearing function is included in order to allow the two
level system to have a finite extension in space.\footnote{%
In turn, a time-window function is included in the interaction to allows the
interaction to occur in a finite time.} Experimental implementations of the
Unruh-DeWitt model have been developed in atomic systems and superconducting
circuits (\cite{ols1}, \cite{ols2} and \cite{sab}), where in the former an
alkali atom as a first quantized system, can serve as a detector for the
second quantized electromagnetic field.

Nevertheless, and despite its great success, the Unruh-DeWitt model cannot
capture the complete interaction between atoms and the electromagnetic field
vacuum. The electromagnetic field is a vector field and carries angular
momentum, but this implies that any study based on the Unruh-DeWitt model
will not capture the anisotropies and orientation dependence of the
entanglement harvesting and will not predict any effect related to the
exchange of the angular momentum of the atoms with the quantum field. In 
\cite{pozas}, a dipole coupling between the electromagnetic field and
hydrogenoid atoms was studied exhaustively. In turn, particle detector
models are ubiquitous as models for experimental setups in quantum optics 
\cite{scull}. The most usual light-matter interaction models, the
Jaynes-Cummings model and its variants, are almost identical to the
Unruh-DeWitt model \cite{scull}, where the rotating wave approximation is
applied and the terms proportional to $\sigma ^{+}a^{\dag }$ and the
Hermitian conjugate are removed from the Hamiltonian. The reason behind this
approximation is that the neglected terms yield bounded oscillations when
integrated in time for the detector-field resonance. These bounded
oscillations can be neglected in the detector-field dynamics compared to the
close-to-resonance rotating wave terms. Removing these terms implies that
microcausality is not guaranteed and that the Hamiltonian is no longer
linear in the field \cite{marti}.

On the other hand, graphene -a monolayer of carbon atoms- has garnered
considerable interest because it is attractive for various electronic and
magnetic applications (\cite{geim}, \cite{fede}, \cite{fede2}). Besides its
novel high-speed electronics properties \cite{das}, graphene is of great
interest from the point of view of fundamental physics as well. The
low-energy electron excitations in graphene are massless Dirac fermions with
a linear energy spectrum (\cite{novo2} and \cite{castro}). This makes
graphene a condensed-matter playground to study various relativistic quantum
phenomena, such as the Klein tunneling and the Casimir effect (\cite{been}
and \cite{fial}). Up to now, most graphene-related studies focused on its
unusual transport properties, but quantum effects arising from interactions
with a quantized electromagnetic field have been neglected.

Recently, quantum electromagnetic field effects have been studied with the
purpose of opening a band gap in the spectrum by illuminating graphene with
circularly polarized light \cite{kib}. In this case the gap appears due to
the formation of composite electron-photon states which are similar to
polaritons in ionic crystals and quantum microcavities (\cite{li}, \cite{low}%
). It should be noted that within the framework of QED, the excitonic
effects can be observed even if real photons are absent and electrons
interact only with vacuum\ fluctuations of the electromagnetic field, by
emitting and reabsorbing virtual photons \cite{bere}. From this, it would be
natural to expect that the photon-induced splitting of the valence and
conductivity bands in\ graphene \cite{kib} will arise due to the vacuum
fluctuations even in the absence of external field pumping. These effects
can be observed by decreasing the effective volume where electron-photon
interactions takes place, which can be accomplished by embedding an electron
system inside a planar microcavity (\cite{liew}).

When the electromagnetic field is coupled to graphene in the long-wavelength
approximation, the minimal coupling $\mathbf{p}\rightarrow \mathbf{p}-e%
\mathbf{A}$ must be applied to the Hamiltonian, which naturally introduces
the Unruh-DeWitt interaction between the detector (in this case, the
sublattice basis) and the quantum field. This allows to study the
entanglement harvesting between two graphene sheets inside a planar
microcavity and in particular, to study the photon-induced splitting of the
valence and conduction bands at small times. From the conceptual viewpoint,
this is a generalization of entanglement harversting to extended or surface
systems and not pointlike systems, such as atomic probes or two-level
systems. It should be stressed that separated electron systems, such as
double-layer graphene, remain strongly coupled by electron-electron
interactions even when they cannot exchange particles, provided that the
layer separation $d$ is comparable to a characteristic distance $l$ between
charge carriers within layers \cite{pla}. One of the consequences of this
remote coupling is a phenomenon called Coulomb drag, in which an electric
current passing through one of the layers causes frictional charge flow in
the other layer and reveals many unpredicted features in double-layer
graphene, such as a larger Coulomb drag when both layers are neutral \cite%
{gor}. Although this phenomenon is considerable for double-layer graphene in
a cavity, when entanglement harvesting -due to the vacuum fluctuations of
the electromagnetic field cavity being studied- time scales much smaller
than the light-crossing time between the layers are considered and the
Coulomb drag can be neglected or, from the point of view of quantum field
theory, we can consider the Coulomb interaction between the layers in the
spacetime region where causality is violated.

Thus, in this work we study the entanglement harvesting between two graphene
sheets inside a cavity, where the monopole detector is given by the natural
interaction of the electrons in graphene with the electromagnetic field. In
particular, the raising and lowering operators that act as the detector are
obtained through the Pauli matrices, which act on the sublattice basis. When
the initial states of electrons are written as eigenstates of the free
Hamiltonian, the effect of the interaction is not trivial because these
eigenstates are written as superpositions of the sublattice basis. In turn,
when the initial states of the electrons are given in a defined sublattice
basis, the entanglement harvesting obtained is identical to that obtained in 
\cite{reznik2} with the main difference coming from the smearing of the
detectors, which in this work are presented by the graphene sheets.

This paper is organized as follows: In Sec. II we introduce the formalism to
compute the time-dependent pertubation theory. In Sec. III we present our 
results and discussions for different initial states of electrons in both
graphene sheets and make a comparison with the Unruh-DeWitt detector. In
Sec. IV we present our conclusions. In Appendices A and B we present a
detailed calculation of the photon propagator and second order contribution
to the reduced quantum operator in time-dependent perturbation theory.

\section{Introduction}

\begin{figure}[tbp]
\centering\includegraphics[width=70mm,height=45mm]{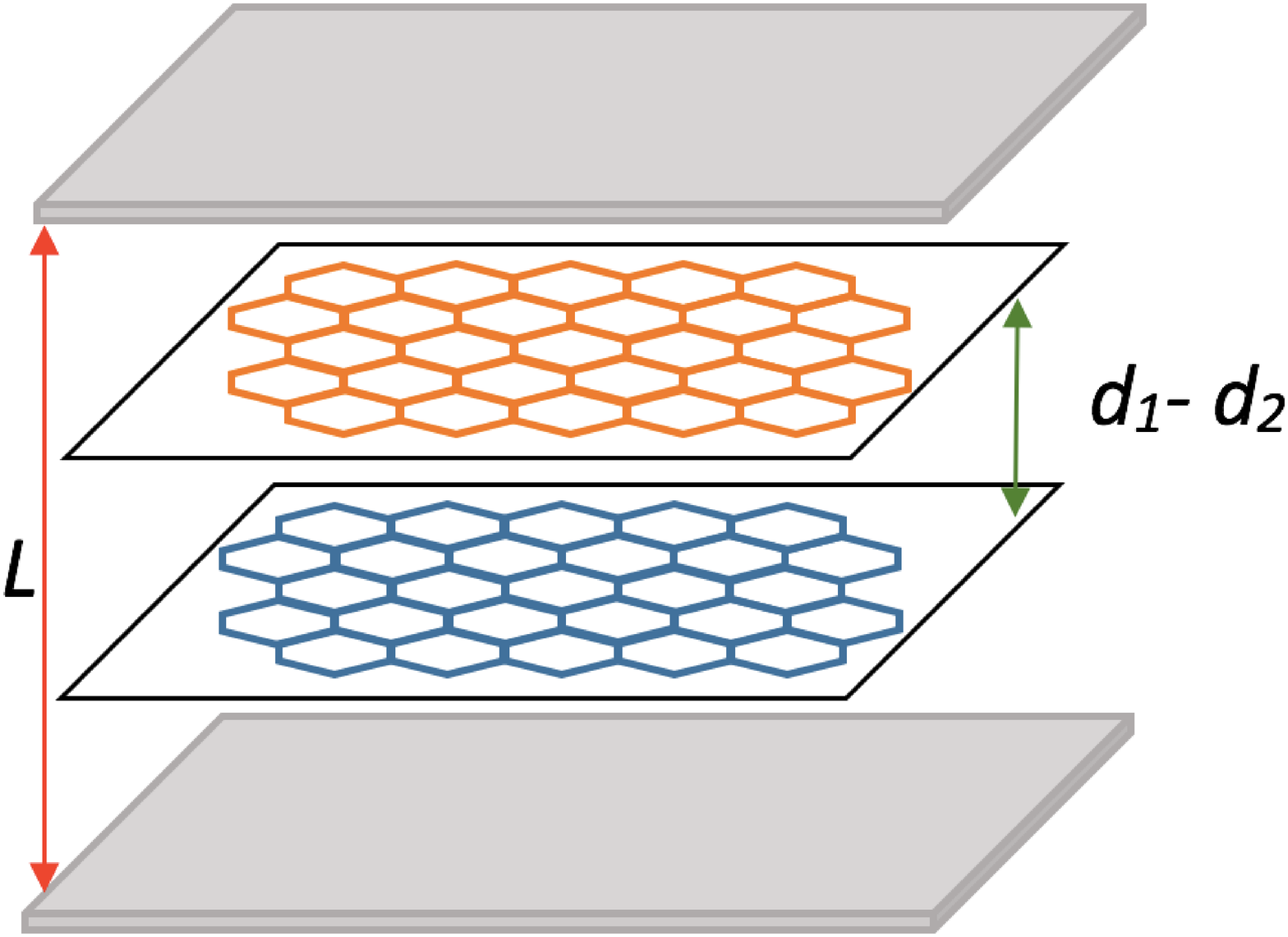}
\caption{The device setup of double-layer graphene inside a planar microcavity.}
\label{doublelayer}
\end{figure}

The Hamiltonian of the double-layer graphene coupled to the electromagnetic
field of the cavity reads (see Fig. \ref{doublelayer})%
\begin{equation}
H=\overset{}{\underset{i=1,2}{\sum }}(v_{F}\mathbf{\sigma }_{i}\mathbf{p}%
_{i}-ev_{F}\mathbf{\sigma }_{i}\mathbf{A}_{i})+H_{F}  \label{1}
\end{equation}%
where $i$ runs over the two electrons, each in different graphene layers%
\footnote{%
The index $i$ should not be confused with the index notation of a vector.},
which can be in either in the valence or conduction band, $\mathbf{A}_{i}$
is the potential vector acting on each electron and $H_{F}$ is the
Hamiltonian of the electromagnetic field $H_{F}=\overset{}{\underset{n,%
\mathbf{q,\lambda }}{\sum }}\hbar \omega _{n,\mathbf{q,\lambda }}a_{n\mathbf{%
q\lambda }}^{\dag }a_{n\mathbf{q\lambda }}$, where $\omega _{n,\mathbf{q}}=c%
\sqrt{q^{2}+(\frac{\pi n}{L})^{2}}$ and $a_{n\mathbf{q\lambda }}$($a_{n%
\mathbf{q\lambda }}^{\dag }$) are the creation and annihilation operators of
the cavity field that obeys the usual commutation relations $\left[ a_{n%
\mathbf{q\lambda }},a_{n^{\prime }\mathbf{q}^{\prime }\mathbf{\lambda }%
^{\prime }}^{\dag }\right] =\delta _{n,n^{\prime }}\delta _{\lambda ,\lambda
^{\prime }}\delta _{\mathbf{qq}^{\prime }}I$ \cite{kib}. 
The quantum electromagnetic field can be written in terms of the creation
and annihilation operators for each mode $\mathbf{q}$ with frequency $\omega
_{n,\mathbf{q}}$ and polarization $\lambda $ as \cite{kib3}%
\begin{equation}
\mathbf{A}_{i}(\mathbf{r},z,t)\mathbf{=\overset{}{\underset{\lambda =\pm ;n,%
\mathbf{q}}{\sum }}}\frac{\gamma }{\sqrt{\omega _{n,\mathbf{q}}}}\mathbf{%
\sin (}\frac{\pi nd_{i}}{L}\mathbf{)(}\widehat{e}_{\mathbf{q}\lambda }a_{n%
\mathbf{q}\lambda }e^{i(\mathbf{q\cdot r-\omega _{n,q}t})}+\widehat{e}_{%
\mathbf{q}\lambda }a_{n\mathbf{q}\lambda }^{\dag }e^{-i(\mathbf{q\cdot
r-\omega _{n,q}t})}\mathbf{)}  \label{2}
\end{equation}%
where $\widehat{e}_{\mathbf{q\lambda }}$ are the polarization directions
orthogonal to the in-plane wave vector of the field $\mathbf{q}$, $L$ is the
distance between the two mirrors of the planar cavity, $S$ is the area of
the graphene sample $\gamma =\sqrt{\frac{\hbar }{\epsilon LS}}$ and $n$ is
the mode index in the $z$ direction. By considering one layer of graphene
placed at $z=d_{1}$ and the second one at $z=d_{2}$, the potential vectors $%
A_{i}$ are given by eq.(\ref{2}) with the replacements $z=d_{1}$ and $z=d_{2}
$ respectively. By considering $H_{0}=\overset{}{\underset{i=1,2}{\sum }}%
v_{F}\mathbf{\sigma }_{i}\mathbf{p}_{i}$ the unperturbed Hamiltonian, a set
of eigenstates can be obtained in terms of the sublattice basis%
\begin{equation}
\left\vert \mathbf{k}_{i}\mathbf{,}s_{i}\right\rangle =\frac{e^{i\mathbf{k}%
_{i}\mathbf{\cdot r}}}{\sqrt{2S}}(\left\vert A_{i}\right\rangle +se^{i\theta
_{i}}\left\vert B_{i}\right\rangle )  \label{3}
\end{equation}%
where $\theta _{i}=\arctan (\frac{k_{y_{i}}}{k_{x_{i}}})$ is the angle of
the wave vector with respect the $x$ axis and $s=\pm 1$ for the conduction
and valence bands.\footnote{%
In the low-wavelength approximation, the wave vector can be approximated at
one of the two inequivalent symmetry points of the Brillouin zone, the $K$
or $K^{\prime }$ valleys. For the sake of simplicity we will consider one
valley.} The elementary electromagnetic field excitations from the vacuum
can be characterized by the wave vector $\mathbf{q}$\ and the helicity,
which can be constructed through the polarization vectors $\widehat{e}_{x}$\
and $\widehat{e}_{y}$\ by redefining $\widehat{e}_{+}=\frac{1}{\sqrt{2}}(%
\widehat{e}_{x}+i\widehat{e}_{y})$\ and $\widehat{e}_{-}=\frac{1}{\sqrt{2}}(%
\widehat{e}_{x}-i\widehat{e}_{y})$\textit{.} In order to express the dot
product $\mathbf{\sigma }_{i}\mathbf{A}_{i}$ we have to consider a
two-dimensional space orthogonal to the $z$ direction. By using the circular
polarization basis, the dot product reads\footnote{%
We are assuming that the virtual photons interacting with the graphene
layers propagate normally with respect to these layers.}%
\begin{equation}
\mathbf{\sigma }_{i}\mathbf{A}_{i}=(\sigma _{x}^{(i)}\widehat{e}_{x}+\sigma
_{y}^{(i)}\widehat{e}_{y})(A_{i}^{+}\widehat{e}_{+}+A_{i}^{-}\widehat{e}%
_{-})=\sqrt{2}\overset{}{\underset{\lambda =\pm }{\sum }}\sigma _{-\lambda
}^{(i)}A_{\lambda }^{(i)}  \label{6}
\end{equation}%
where $\lambda =\pm 1$ for both helicities, $\sigma _{\lambda }^{(i)}=\frac{1%
}{2}(\sigma _{x}^{(i)}+\lambda i\sigma _{y}^{(i)})$ and where 
\begin{equation}
A_{\lambda }^{(i)}=\overset{}{\underset{n,\mathbf{q}}{\sum }}\frac{\gamma 
\mathbf{\sin (}\frac{\pi nd_{i}}{L}\mathbf{)}}{\sqrt{\omega _{n,\mathbf{q}}}}%
(a_{n\mathbf{q}\lambda }e^{i(\mathbf{q\cdot r-\omega _{n,q}t})}+a_{n\mathbf{q%
}\lambda }^{\dag }e^{-i(\mathbf{q\cdot r-\omega _{n,q}t})})  \label{7}
\end{equation}%
In order to compute the coupling between the valence and conduction bands
with the circular polarized photons, the following relations must be taken
into account $\sigma _{+}\left\vert \mathbf{k,}A\right\rangle =0$, $\sigma
_{+}\left\vert \mathbf{k,}B\right\rangle =\left\vert \mathbf{k,}%
A\right\rangle $, $\sigma _{-}\left\vert \mathbf{k,}A\right\rangle
=\left\vert \mathbf{k,}B\right\rangle $, $\sigma _{-}\left\vert \mathbf{k,}%
B\right\rangle =0$, $\sigma _{+}\left\vert \mathbf{k,+}\right\rangle =\frac{1%
}{2}e^{i\eta _{\mathbf{k}}}\left( \left\vert \mathbf{k,+}\right\rangle
+\left\vert \mathbf{k,-}\right\rangle \right) $, $\sigma _{-}\left\vert 
\mathbf{k,+}\right\rangle =\frac{1}{2}e^{-i\eta _{\mathbf{k}}}\left(
\left\vert \mathbf{k,+}\right\rangle -\left\vert \mathbf{k,-}\right\rangle
\right) $, $\sigma _{+}\left\vert \mathbf{k,-}\right\rangle =-\frac{1}{2}%
e^{i\eta _{\mathbf{k}}}\left( \left\vert \mathbf{k,+}\right\rangle
+\left\vert \mathbf{k,-}\right\rangle \right) $ and $\sigma _{-}\left\vert 
\mathbf{k,-}\right\rangle =\frac{1}{2}e^{-i\eta _{\mathbf{k}}}\left(
\left\vert \mathbf{k,+}\right\rangle -\left\vert \mathbf{k,-}\right\rangle
\right) $ (see \cite{kib}). From these relations we can see that this model
is similar to those used in entanglement harvesting from two detectors \cite%
{pozas}, where $\sigma _{\pm }$ are the detector's energy raising and
lowering operators. In this work, this two-level system is the sublattice
basis, which implies that one photon with a definitive helicity is absorbed
whenever a delocalized electron jumps from the $A$ sublattice to $B$
sublattice or a photon is emitted when an electron jumps from the $B$
sublattice to the $A$ sublattice. But the stationary states in graphene are
given by the eigenvectors of the Hamiltonian which can be written as
superpositions in the sublattice basis (see eq.(\ref{3})). This implies that
the entanglement harvesting of the two graphene layers is more subtle
because the absorbtion and emission of virtual photons imply a superposition
of valence and conduction bands with definite incoming and outgoing
momentum. In turn, the system under study is a generalization of pointlike
systems, where the monopole detectors raise and lower the two discrete
energy levels. In the case of double-layer graphene, the detector is given
by the interaction $\mathbf{\sigma A}$, where $\mathbf{A}$ is now evaluated
in each graphene layer. The entanglement harvesting on surfaces implies, at
least two energy bands, and the possible transitions are ruled by the energy
conservation given by the momentum of the electrons in both graphene sheets.
In the literature, entanglement harvesting is investigated using a pointlike
approximation for the detector model, which has no extension and interacts
with the field only at the spacetime point where it is placed. This
assumption, which can be considered an approximation for real detectors with
finite size, results in ultraviolet divergences. Several regularization
schemes yield different transition probabilities \cite{sch}. In the case of
double-layer graphene, this problem is not present due to the natural
spatial smearing of the interaction between the electromagnetic field and
graphene sheets.

In order to compute the entanglement of electrons we can work perturbatively
to second order in the interacting Hamiltonian $H_{int}=\sqrt{2}\overset{}{%
\underset{\lambda =\pm }{\sum }}\sigma _{-\lambda }^{(i)}A_{\lambda }^{(i)}$%
, where the interaction picture time evolution operator $U$ for the full
system is 
\begin{equation}
U=U^{(0)}+U^{(1)}+U^{(2)}+...  \label{8a}
\end{equation}%
where $U^{(0)}=I$, $U^{(1)}=-i\int_{-\infty }^{t}dt^{\prime
}H_{int}(t^{\prime })$, $U^{(2)}=-\int_{-\infty }^{t}dt^{\prime
}H_{int}(t^{\prime })\int_{-\infty }^{t^{\prime }}dt^{\prime \prime
}H_{int}(t^{\prime \prime })$ and $%
H_{int}(t)=e^{-i(H_{0}+H_{F})t}V(t)e^{i(H_{0}+H_{F})t}$. Then, given an
initial density matrix $\rho _{0}$, the final density matrix $\rho _{T}$ is
hence given by%
\begin{equation}
\rho _{T}=U\rho _{0}U^{\dag }=[I+U^{(1)}+U^{(2)}+...]\rho
_{0}[I+U^{(1)}+U^{(2)}+...]  \label{a9}
\end{equation}%
If we write $\rho _{T}=\rho _{0}+\rho _{T}^{(1)}+\rho _{T}^{(2)}+...$, then%
\begin{gather}
\rho _{T}^{(1)}=U^{(1)}\rho _{0}+\rho _{0}U^{(1)\dag }  \label{a10} \\
\rho _{T}^{(2)}=U^{(1)}\rho _{0}U^{(1)\dag }+U^{(2)}\rho _{0}+\rho
_{0}U^{(2)\dag }  \notag
\end{gather}%
In order to rearrange the notation, we can write $\rho
_{T}^{(i,j)}=U^{(i)}\rho _{0}U^{(j)\dag }$ and therefore, the time-evolved
density matrix can be written as a sum of terms of the form $\rho =\rho
_{0}+\rho ^{(1,0)}$\thinspace $+\rho ^{(0,1)}+\rho ^{(2,0)}+\rho
^{(0,2)}+\rho ^{(1,1)}+...$. Because we are going to analyze entanglement
and correlation harvesting of both graphene layers from the vacuum
fluctuations of the quantum electromagnetic field, we can consider that the
initial state of the electron-electron quantum field system is 
\begin{equation}
\rho _{0}=\left\vert \Omega _{0}\right\rangle \left\langle \Omega
_{0}\right\vert \otimes \rho _{G}  \label{a11}
\end{equation}%
where $\left\vert \Omega _{0}\right\rangle =\left\vert \Omega
_{0}^{(+)},\Omega _{0}^{(-)}\right\rangle $ is the vacuum state of the
electromagnetic field with circular polarization $\pm ~$and $\rho _{G}$ is
the initial density matrix of the electron-electron system, where without
loss of generality, we can consider that both electrons are in the
conduction band with momenta $\mathbf{k_{1}}$ and $\mathbf{k_{2}}$
respectively, or both electrons are in the sublattice basis $A$ with momenta 
$\mathbf{k_{1}}$ and $\mathbf{k_{2}}$ respectively. We are interested in the
partial state of the electrons in the graphene sheet after the interaction
with the quantum field, which is given by 
\begin{equation}
\rho (t)=Tr_{A}(U\rho _{0}U^{\dag })  \label{a12}
\end{equation}%
This means that the nondiagonal terms in the field produced by time
evolution will be not be relevant for our purposes. In particular, any
contribution for which the parities of $i$ and $j$ are different will give a
zero contribution to the electrons in graphene final states as long as the
initial state of the field is diagonal in the Fock basis, which is the case
for the vacuum or any incoherent superposition of Fock states such as a
thermal state. Then, the unique term to be computed is $\rho _{T}^{(2)}$ and
the trace over the field basis must be carried out. The $Tr_{A}(U^{(2)}\rho
_{0})=-\frac{1}{2}\int_{-\infty }^{t}\int_{-\infty }^{t}dt_{1}dt_{2}Tr_{A}%
\left[ H_{int}(t_{1})H_{int}(t_{2})\rho _{0}\right] $ contribution in eq.(%
\ref{a10}) can be written as%
\begin{equation}
Tr_{\phi }(U^{(2)}\rho _{0})=-(ev_{F})^{2}\overset{}{\underset{%
i,j=1,2;\lambda ,\lambda ^{\prime }}{\sum }}\int d^{2}\mathbf{r_{1}}\int
d^{2}\mathbf{r_{2}}\int_{-\infty }^{t}\int_{-\infty }^{t}dt_{1}dt_{2}\Delta
_{\lambda ,\lambda ^{\prime }}^{(i,j)}(\mathbf{r}_{1},t_{1},\mathbf{r}%
_{2},t_{2})\sigma _{-\lambda }^{(i)}(t_{1})\sigma _{-\lambda ^{\prime
}}^{(j)}(t_{2})\rho _{G}  \label{14}
\end{equation}%
where%
\begin{gather}
\Delta _{\lambda ,\lambda ^{\prime }}^{(i,j)}(\mathbf{r}_{1},t_{1},\mathbf{r}%
_{2},t_{2})=\left\langle \Omega _{0}\right\vert A_{\lambda
}^{(i)}(t_{1})A_{\lambda ^{\prime }}^{(j)}(t_{2})\left\vert \Omega
_{0}\right\rangle =  \label{15} \\
\delta _{\lambda \lambda ^{\prime }}\overset{}{\underset{n,\mathbf{q}}{\sum }%
}\frac{\gamma ^{2}}{\omega _{n,\mathbf{q}}}\mathbf{\sin (}\frac{\pi nd_{i}}{L%
}\mathbf{)\sin (}\frac{\pi nd_{j}}{L}\mathbf{)}e^{i(\mathbf{q\cdot
r_{1}-\omega _{n,q}t_{1}})}e^{-i(\mathbf{q\cdot r_{2}-\omega _{n,q}t_{2}})} 
\notag
\end{gather}%
is the photon propagator, where we have used that $A_{\lambda }^{(i)}(%
\mathbf{r},t)=e^{\frac{i}{\hbar }H_{0F}t}A_{\lambda }^{(i)}e^{-\frac{i}{%
\hbar }H_{0F}t}$ and where $\sigma _{-\lambda }^{(i)}(t)=e^{\frac{i}{\hbar }%
H_{0S}t}\sigma _{-\lambda }^{(i)}e^{-\frac{i}{\hbar }H_{0S}t}$. The photon
propagator can be computed exactly (see Appendix A) and the result reads $%
\Delta _{\lambda ,\lambda ^{\prime }}^{(i,j)}(\Delta t,\left\vert \Delta 
\mathbf{r}\right\vert )=\delta _{\lambda \lambda ^{\prime
}}F_{ij}(\left\vert x\right\vert )$ where%
\begin{equation}
F_{ij}(\left\vert x\right\vert )=\frac{\gamma ^{2}\sin (\frac{\pi d_{i}}{L}%
)\sin (\frac{\pi d_{j}}{L})\sinh (\frac{\pi \left\vert x\right\vert }{L})}{%
16\pi \left\vert x\right\vert \sin (\frac{\pi (d_{i}-d_{j}-i\left\vert
x\right\vert )}{2L})\sin (\frac{\pi (d_{i}+d_{j}-i\left\vert x\right\vert )}{%
2L})\sin (\frac{\pi (d_{i}-d_{j}+i\left\vert x\right\vert )}{2L})\sin (\frac{%
\pi (d_{i}+d_{j}+i\left\vert x\right\vert )}{2L})}  \label{16}
\end{equation}%
where $\left\vert x\right\vert =\sqrt{c^{2}\Delta t^{2}-\left\vert \Delta 
\mathbf{r}\right\vert ^{2}}$, with $\Delta t=t_{1}-t_{2}$ and $\Delta 
\mathbf{r=r_{1}-r_{2}}$. In the last equation, the infinite sum of modes has
been carried out, although it is known that realistic cavities are not good
cavities for the whole frequency spectrum, thus an improved version of the
model introduced in this work should introduce a mode cutoff. Nevertheless,
this cutoff would imply that the usual light-matter interaction violates
causality. Then, although the model is ideal and does not represent real
cavities, it is consistent with causality. In a similar way, the other two
contributions to $\rho $ at second order read%
\begin{equation}
Tr_{\phi }(\rho _{0}U^{(2)\dag })=-(ev_{F})^{2}\overset{}{\underset{%
i,j=1,2;\lambda ,\lambda ^{\prime }}{\sum }}\int d^{2}\mathbf{r_{1}}\int
d^{2}\mathbf{r_{2}}\int_{-\infty }^{t}\int_{-\infty }^{t}dt_{1}dt_{2}\Delta
_{\lambda ,\lambda ^{\prime }}^{(ij)}(\mathbf{r}_{1},t_{1},\mathbf{r}%
_{2},t_{2})\rho _{G}\sigma _{-\lambda }^{(i)\dag }(t_{1})\sigma _{-\lambda
^{\prime }}^{(j)\dag }(t_{2})  \label{17}
\end{equation}%
and%
\begin{equation}
Tr_{\phi }(U^{(1)}\rho _{0}U^{(1)\dag })=2(ev_{F})^{2}\overset{}{\underset{%
i,j=1,2;\lambda ,\lambda ^{\prime }}{\sum }}\int_{-\infty }^{t}\int_{-\infty
}^{t}dt_{1}dt_{2}\Delta _{\lambda ,\lambda ^{\prime }}^{(i,j)\ast }(\mathbf{r%
}_{1},t_{1},\mathbf{r}_{2},t_{2})\sigma _{-\lambda }^{(i)}(t_{1})\rho
_{G}\sigma _{-\lambda }^{(j)\dag }(t_{2})  \label{18a}
\end{equation}%
Collecting all the terms, the reduced state reads%
\begin{gather}
\rho =Tr_{\phi }(\rho (t))=-(ev_{F})^{2}\overset{}{\underset{i,j=1,2;\lambda 
}{\sum }}\int_{-\infty }^{t}\int_{-\infty }^{t}dt_{1}dt_{2}\Delta _{\lambda
,\lambda ^{\prime }}^{(ij)}(\mathbf{r}_{1},t_{1},\mathbf{r}_{2},t_{2})\times 
\label{19} \\
\left[ \sigma _{-\lambda }^{(i)}(t_{1})\sigma _{-\lambda }^{(j)}(t_{2})\rho
_{G}+\rho _{G}\sigma _{-\lambda }^{(i)\dag }(t_{1})\sigma _{-\lambda
}^{(j)\dag }(t_{2})-2\sigma _{-\lambda }^{(i)}(t_{1})\rho _{G}\sigma
_{-\lambda }^{(j)\dag }(t_{2})\right]   \notag
\end{gather}%
\begin{figure}[tbp]
\centering\includegraphics[width=70mm,height=55mm]{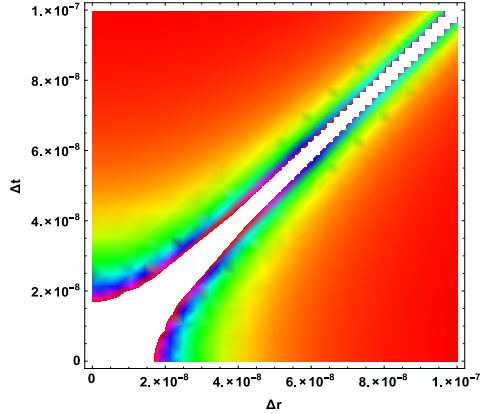}
\caption{Photon cavity propagator as a function of space and time.}
\label{propa}
\end{figure}
In Fig. \ref{propa} the photon propagator in the cavity is shown as a
function of $\left\vert x\right\vert $ for $d_{1}/L=0.4$ and $d_{2}/L=0.6$.
As it can be seen, the propagator does not vanish outside the light cone,
which implies the emergence of correlations between the two graphene sheets
at $t<c/\left\vert d_{2}-d_{1}\right\vert $. This implies the generation of
a correlated state from an uncorrelated one only by local interactions
because the field vacuum is an entangled state between spacelike separated
regions. In turn, the nonzero probability of an electron in the graphene
sheet to get excited outside the light cone is independent of the remaining
electron in the other graphene sheet and thus no information is carried over
spacelike distance. The main difference between the result obtained for $%
\rho $ in double-layer graphene and the pointlike detectors is the spatial
integration over the constrained space in which the electrons can move. When
real detectors are modeled, a smeared function must be introduced in the
interaction which introduces the spatial integration (see \cite{mon}). Both
electrons are delocalized in each graphene sheet and can become entangled by
merely letting them interact with the field vacuum state. The system becomes
entangled because they swap entanglement from the vacuum rather than by
interacting through the exchange of real field quanta. \qquad 

Finally, the matrix elements $\left\langle \mathbf{k}_{1}^{\prime
},s_{1}^{\prime }\mathbf{,k}_{2}^{\prime },s_{2}^{\prime }\right\vert
Tr_{\phi }(U^{(2)}\rho _{0})\left\vert \mathbf{k}_{1},s_{1}\mathbf{,k}%
_{2},s_{2}\right\rangle $ read (see Appendix B)%
\begin{gather}
\left\langle \mathbf{k}_{1}^{\prime },s_{1}^{\prime }\mathbf{,k}_{2}^{\prime
},s_{2}^{\prime }\right\vert \rho (t)\left\vert \mathbf{k}_{1},s_{1}\mathbf{%
,k}_{2},s_{2}\right\rangle =-(ev_{F})^{2}\delta (\mathbf{k_{1}-k_{1}^{\prime
}+k_{2}-k_{2}^{\prime }})\overset{}{\underset{i,j=1,2;\lambda }{\sum }}%
\int_{0}^{t}\int_{0}^{t}dt_{1}dt_{2}\mathcal{F}_{ij}\left( \mathbf{%
k_{2}-k_{2}^{\prime },}\Delta t\right) \times   \label{20} \\
\left\langle s_{1}^{\prime },s_{2}^{\prime }\right\vert \left( \sigma
_{-\lambda }^{(i)}(t_{1})\sigma _{-\lambda }^{(j)}(t_{2})\rho _{G}-2\sigma
_{-\lambda }^{(i)}(t_{1})\rho _{G}\sigma _{-\lambda }^{(j)\dag }(t_{2})+\rho
_{G}\sigma _{-\lambda }^{(i)\dag }(t_{1})\sigma _{-\lambda }^{(j)\dag
}(t_{2})\right) \left\vert s_{1},s_{2}\right\rangle   \notag
\end{gather}%
where (see eq.(\ref{ap7.0})) reads%
\begin{equation}
\mathcal{F}_{ij}\left( \mathbf{k_{2}-k_{2}^{\prime },}\Delta t\right) =\int
d^{2}\mathbf{\Delta r}e^{-i(\mathbf{k_{2}-k_{2}^{\prime })\cdot \Delta r}%
}F_{ij}(\sqrt{\Delta t^{2}-\left\vert \Delta \mathbf{r}\right\vert ^{2}})
\label{21}
\end{equation}%
and $\left\vert s_{1},s_{2}\right\rangle =\left\vert s_{1}\right\rangle
\otimes \left\vert s_{2}\right\rangle $ is an arbitrary basis, for example
the sublattice basis, in which case $s_{i}=A,B$ or the valence-conduction
band basis, in which case $s_{i}=\pm $. The Dirac delta $\delta (\mathbf{%
k_{1}-k_{1}^{\prime }+k_{2}-k_{2}^{\prime }})$ implies momentum conservation
and $\mathbf{k_{i}}$ $\mathbf{(k_{i}^{\prime })}$ are the initial(final)
momentum of both electrons.

\section{Results and discussions}

In order to obtain the critical parameters in which the reduced quantum
state is entangled, we can expand $Tr_{\phi }(\rho (t))$ in small values of $%
t$ in Eq.(\ref{19})

\begin{gather}
\left\langle \mathbf{k}_{1}^{\prime },s_{1}^{\prime }\mathbf{,k}_{2}^{\prime
},s_{2}^{\prime }\right\vert \rho (t)\left\vert \mathbf{k}_{1},s_{1}\mathbf{%
,k}_{2},s_{2}\right\rangle =-(ev_{F}t)^{2}\delta (\mathbf{%
k_{1}-k_{1}^{\prime }+k_{2}-k_{2}^{\prime }})\times   \label{r1} \\
\overset{}{\underset{i,j=1,2;\lambda }{\sum }}\mathcal{F}_{ij}\left( \mathbf{%
k_{2}-k_{2}^{\prime },}0\right) \times \left\langle s_{1}^{\prime
},s_{2}^{\prime }\right\vert \left( \sigma _{-\lambda }^{(i)}\sigma
_{-\lambda }^{(j)}\rho _{G}-2\sigma _{-\lambda }^{(i)}\rho _{G}\sigma
_{\lambda }^{(j)}+\rho _{G}\sigma _{\lambda }^{(i)}\sigma _{\lambda
}^{(j)}\right) \left\vert s_{1},s_{2}\right\rangle   \notag
\end{gather}%
where we have used that $\sigma _{-\lambda }^{(i)\dag }=\sigma _{\lambda
}^{(i)}$. Considering as initial state $\rho _{G}=\left\vert A\mathbf{,}%
A\right\rangle \left\langle A\mathbf{,}A\right\vert $ where both electrons
in each graphene sheet have nonzero amplitude in the $A$ sublattice basis,
the normalized reduced quantum state can be written in the basis $\left\vert
A,A\right\rangle $, $\left\vert A,B\right\rangle $, $\left\vert
B,A\right\rangle $ and $\left\vert B,B\right\rangle $ as\qquad 
\begin{equation}
\rho (t)=\left[ 
\begin{array}{cccc}
1-2(ev_{F})^{2}t^{2}\left[ \mathcal{F}_{11}+\mathcal{F}_{22}\right]  & 0 & 0
& 2(ev_{F})^{2}t^{2}\mathcal{F}_{12} \\ 
0 & 2(ev_{F})^{2}t^{2}\mathcal{F}_{22} & -2(ev_{F})^{2}t^{2}\mathcal{F}_{12}
& 0 \\ 
0 & -2(ev_{F})^{2}t^{2}\mathcal{F}_{12} & 2(ev_{F})^{2}t^{2}\mathcal{F}_{11}
& 0 \\ 
2(ev_{F})^{2}t^{2}\mathcal{F}_{12} & 0 & 0 & 0%
\end{array}%
\right]   \label{r2}
\end{equation}%
where $\mathcal{F}_{ij}$ is a function of $\mathbf{k_{2}-k_{2}^{\prime }}$
and momentum conservation is understood. This density matrix has the form of
the so-called $X$ state \cite{ali} and is positive at leading order in $%
O((\gamma ev_{F})^{2})$ and all the perturbative corrections of $\rho $ to
the final density matrix are traceless. Therefore, the trace of the final
state of $\rho $ is always preserved, independent of up to which order $O(n)$
in the coupling constant the corrections are taken into account.

The $X$ states are those in which several matrix elements are zero ($\rho
_{12}=\rho _{13}=\rho _{24}=\rho _{34}=0$) \cite{tuyu}. In turn, many
well-known and useful families of states have an $X$ form, including the
Bell states, Werner states \cite{wer}, and isotropic states \cite{tuyu}.
Recently, it was shown numerically that all two-qubit mixed states are
equivalent to $X$ states by a single entanglement-preserving unitary
transformation, so concurrence and other entanglement measures of such an $X$
state are equal to those of the original general state \cite{men}. In
general, a density matrix is said to be inseparable or entangled if it
cannot be expressed as a convex sum of local density matrices \cite{wer}. In
the present case of a 2$\times $ 2 system, a necessary and sufficient
condition for inseparability is that the negativity be positive, where the
negativity $\mathcal{N}$ is defined as the lowest eigenvalue of the partial
transpose of $\rho $ (\cite{per}, \cite{horo} and \cite{vid}). The
negativity is an entanglement monotone that for two-qubit settings only
vanishes for separable states and is defined as%
\begin{equation}
\mathcal{N}(\rho )=\overset{}{\underset{\alpha \in \sigma \left[ \rho
^{\Gamma _{2}}\right] }{\sum }}\frac{\left\vert \alpha _{i}\right\vert
-\alpha _{i}}{2}  \label{r3.1}
\end{equation}%
where $\alpha _{i}$ are the eigenvalues of the partial transpose of $\rho
^{\Gamma _{2}}=(I\otimes T)\rho $ with respect to the second system. This
partial transpose reads%
\begin{equation}
\rho ^{\Gamma _{2}}=(I\otimes T)\rho =\left[ 
\begin{array}{cccc}
1-2(ev_{F}t)^{2}\left[ \mathcal{F}_{11}+\mathcal{F}_{22}\right]  & 0 & 0 & 
-2(ev_{F}t)^{2}\mathcal{F}_{12} \\ 
0 & 2(ev_{F}t)^{2}\mathcal{F}_{22} & 2(ev_{F}t)^{2}\mathcal{F}_{12} & 0 \\ 
0 & 2(ev_{F}t)^{2}\mathcal{F}_{12} & 2(ev_{F}t)^{2}\mathcal{F}_{11} & 0 \\ 
-2(ev_{F}t)^{2}\mathcal{F}_{12} & 0 & 0 & 0%
\end{array}%
\right]   \label{r4}
\end{equation}%
and the eigenvalues read%
\begin{gather}
\alpha _{1}=\frac{1}{2}-e^{2}v_{F}^{2}t^{2}(\mathcal{F}_{11}+\mathcal{F}%
_{22})+\frac{1}{2}\sqrt{4e^{4}t^{4}v_{F}^{4}\left[ (\mathcal{F}_{11}+%
\mathcal{F}_{22})^{2}+4\mathcal{F}_{12}^{2}\right] -4e^{2}t^{2}v_{F}^{2}(%
\mathcal{F}_{11}+\mathcal{F}_{22})+1}  \label{r4.1} \\
\alpha _{2}=\frac{1}{2}-e^{2}v_{F}^{2}t^{2}(\mathcal{F}_{11}+\mathcal{F}%
_{22})-\frac{1}{2}\sqrt{4e^{4}t^{4}v_{F}^{4}\left[ (\mathcal{F}_{11}+%
\mathcal{F}_{22})^{2}+4\mathcal{F}_{12}^{2}\right] -4e^{2}t^{2}v_{F}^{2}(%
\mathcal{F}_{11}+\mathcal{F}_{22})+1}  \notag \\
\alpha _{3}=e^{2}v_{F}^{2}t^{2}\left( \mathcal{F}_{11}+\mathcal{F}_{22}+%
\sqrt{(\mathcal{F}_{11}-\mathcal{F}_{22})^{2}+4\mathcal{F}_{12}^{2}}\right) 
\notag \\
\alpha _{4}=e^{2}v_{F}^{2}t^{2}\left( \mathcal{F}_{11}+\mathcal{F}_{22}-%
\sqrt{(\mathcal{F}_{11}-\mathcal{F}_{22})^{2}+4\mathcal{F}_{12}^{2}}\right) 
\notag
\end{gather}

\begin{figure}[tbh]
\begin{minipage}{0.48\linewidth}
\includegraphics[width=70mm,height=55mm]{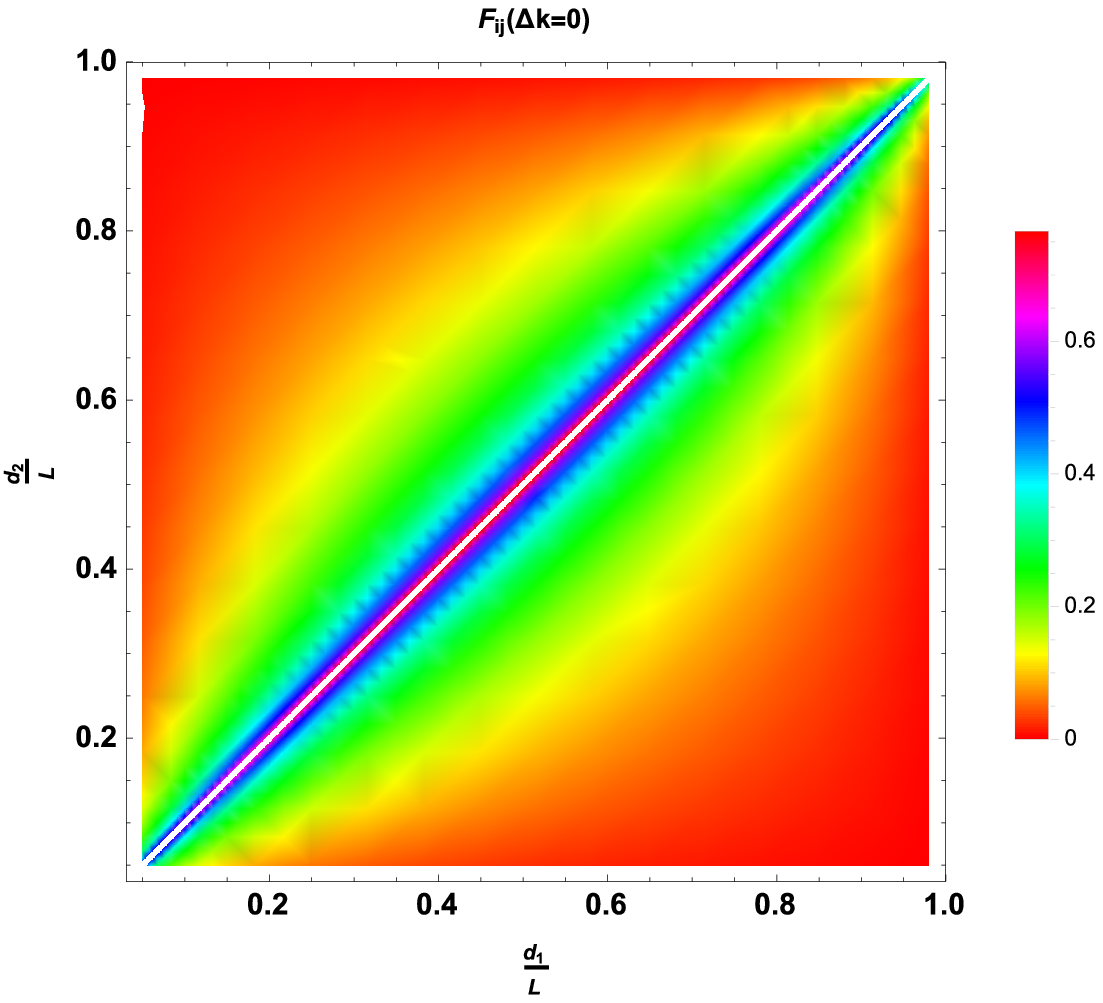}
\caption{The function $\mathcal{F}_{ij}$ as a function of the relative distance
of both graphene sheets with respect to the cavity. }
\label{fij}
\end{minipage}
\hspace{0.07cm} 
\begin{minipage}{0.5\linewidth}
\centering
\includegraphics[width=70mm,height=55mm]{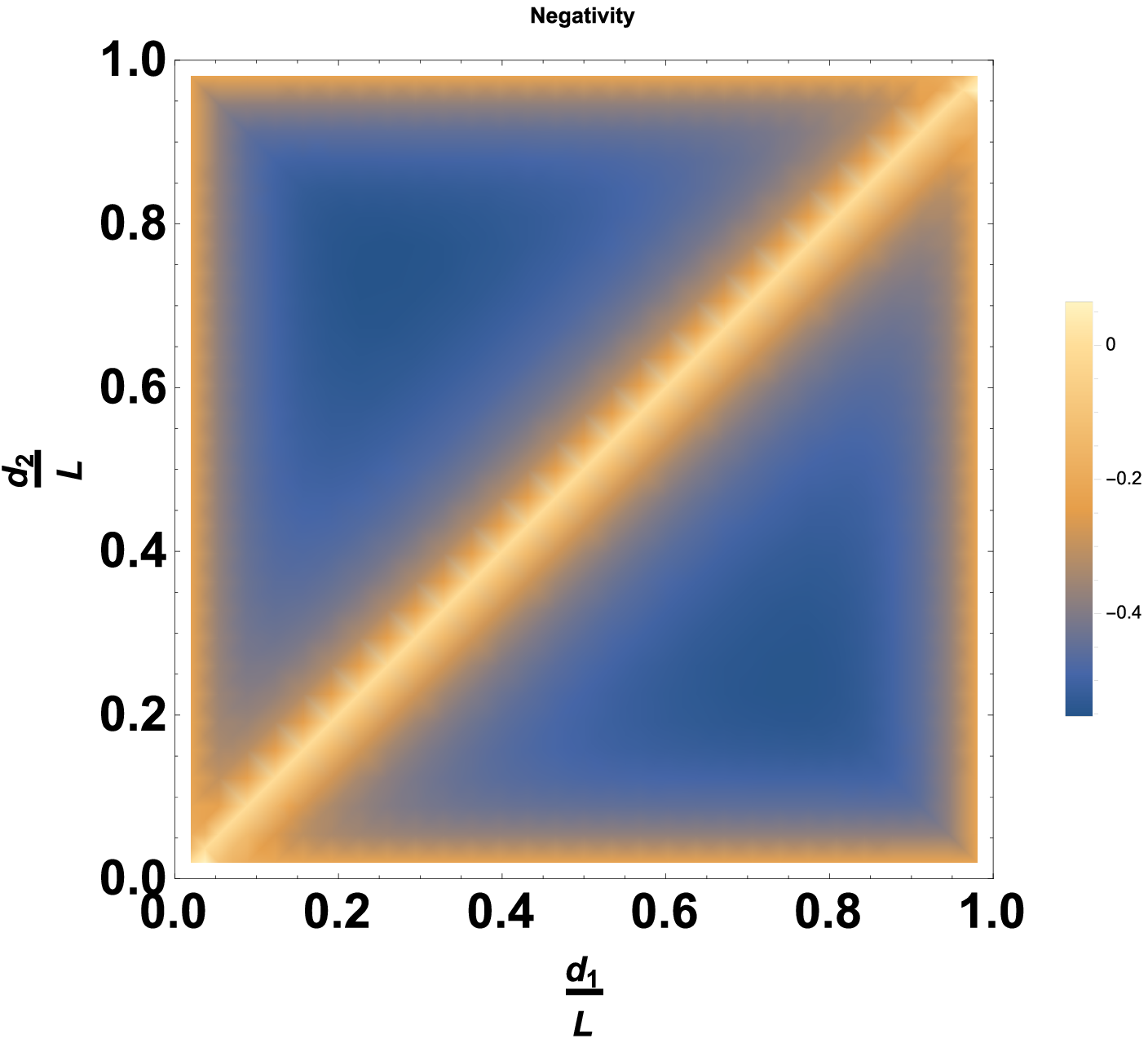}
\caption{The negativity measure as a function of each graphene layer relative
distance with respect the cavity.}
\label{negativity1}
\end{minipage}
\end{figure}

The first two eigenvalues cannot be negative because this would imply that $%
16e^{4}t^{4}v_{F}^{4}\mathcal{F}_{12}^{2}<0$. The only eigenvalue that can
be negative is $\alpha _{4}$. We shall therefore use the negativity as a
measure of entanglement. The following expression is obtained for the
negativity%
\begin{equation}
\mathcal{N}=e^{2}v_{F}^{2}t^{2}\left( \sqrt{(\mathcal{F}_{11}-\mathcal{F}%
_{22})^{2}+4\mathcal{F}_{12}^{2}}-\mathcal{F}_{11}-\mathcal{F}_{22}\right) 
\label{r5}
\end{equation}

\begin{figure}[tbp]
\centering\includegraphics[width=60mm,height=30mm]{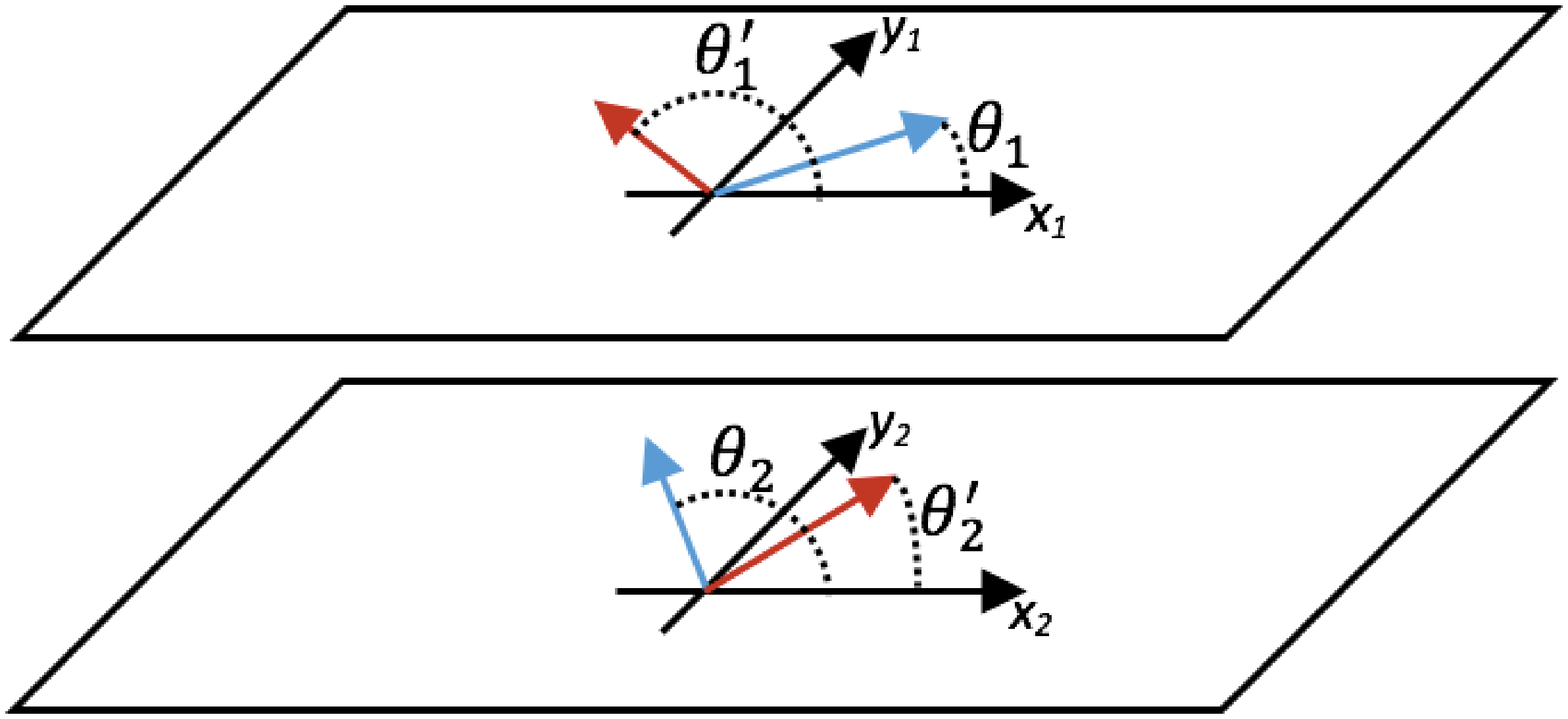}
\caption{Initial and final angles of both conduction electrons in each
graphene sheet. }
\label{angles}
\end{figure}
The last equation is the sum of a local term $\mathcal{F}_{11}+\mathcal{F}%
_{22}$ that depends on the properties of just one of the graphene sheets and
a nonlocal term $\sqrt{(\mathcal{F}_{11}-\mathcal{F}_{22})^{2}+4\mathcal{F}%
_{12}^{2}}$ that depends on the properties of both graphene sheets. This
implies a direct competition between nonlocal, entangling exchange and local
noise, which implies that in order to have entanglement between the graphene
sheets, the nonlocal term must overcome the single-graphene sheet noise
excitations, as it occurs with atoms \cite{resnik}. For the set of values of 
$d_{1}/L$, $d_{2}/L$ and $\Delta k\mathbf{=\left\vert k_{2}-k_{2}^{\prime
}\right\vert }$ in which $\mathcal{N}$ is positive, the double-layer
graphene becomes entangled for times smaller than the light crossing time $t<%
\frac{\left\vert d_{1}-d_{2}\right\vert }{c}$. In order to obtain analytical
results in the case where $\Delta k=0$, instead of computing the sum over $n$
as done in\ Appendix A, we can compute the integral over $\Delta \mathbf{r}$%
. Then $\mathcal{F}_{ij}$ can be written with the sum over $n$%
\begin{equation}
\mathcal{F}_{ij}(\Delta k)=\frac{\gamma ^{2}}{16\pi ^{2}}\overset{\infty }{%
\underset{n=1}{\sum }}\frac{\sin (\frac{n\pi d_{i}}{L})\sin (\frac{n\pi d_{j}%
}{L})}{\sqrt{\Delta k^{2}+\frac{n^{2}\pi ^{2}}{L^{2}}}}  \label{r6}
\end{equation}%
which for the case where $\Delta k=0$ reads%
\begin{gather}
\mathcal{F}_{ij}(k)=\frac{L}{\pi }\ln \left[ \frac{e^{-i(d_{i}+d_{j})\frac{%
\pi }{L}}-1}{e^{id_{i}\frac{\pi }{L}}-e^{id_{j}\frac{\pi }{L}}}\right] 
\label{r7} \\
+\frac{L^{3}\Delta k^{2}}{4\pi ^{3}}\left[ Li_{3}(e^{-i(d_{1}-d_{2})\frac{%
\pi }{L}})+Li_{3}(e^{i(d_{1}-d_{2})\frac{\pi }{L}})-Li_{3}(e^{-i(d_{1}+d_{2})%
\frac{\pi }{L}})-Li_{3}(e^{i(d_{1}+d_{2})\frac{\pi }{L}})\right] +O(\Delta
k^{4})  \notag
\end{gather}%
where we have expanded $\frac{1}{\sqrt{k^{2}+\frac{n^{2}\pi ^{2}}{L^{2}}}}%
\sim \frac{L}{n\pi }-\frac{k^{2}}{2}(\frac{L}{n\pi })^{3}+O(k^{4})$. In Fig. %
\ref{fij} and Fig. \ref{negativity1} the function $\mathcal{F}_{ij}$ and the
negativity are shown as  functions of the dimensionless parameters $d_{1}/L$
and $d_{2}/L$ for $\Delta \mathbf{k=0}$. As expected, the negativity is
larger when the layer separation is smaller at lowest order in $t$. An
electron in one graphene layer has a nonzero probability of getting excited
outside the light cone, but this probability is completely independent of
the electron in the other graphene layer, so no information is being carried
over a spacelike distance.

\begin{figure}[tbp]
\centering\includegraphics[width=100mm,height=60mm]{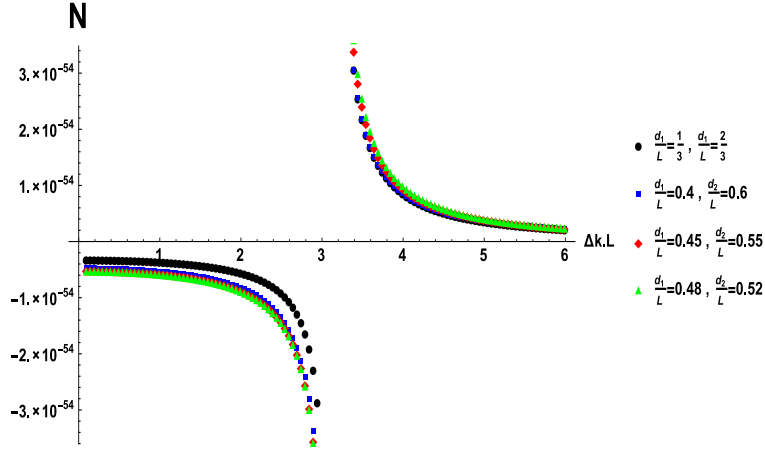}
\caption{The negativity as a function of the momentum transfer for different
relative distances $d_{1}/L$ and $d_{2}/L$. }
\label{trak1}
\end{figure}
In turn, by numerically computing the integral in Eq.(\ref{21}) numerically
for $\Delta t=0$ and $\Delta k\neq 0$ for different sets of values of $%
d_{1}/L$ and $d_{2}/L$, the negativity estimator shows a critical value of $%
\Delta kL\sim 3.2$ where the negativity changes sign (see figure \ref{trak1}
). By considering $L=500nm$ as a normal microcavity, the induced gap is $%
\epsilon _{G}\sim 6\times 10^{-15}$eV which is smaller than typical induced
gaps in normal semiconductors \cite{kib2}. Following the same procedure, we
can consider that the initial quantum state for the two electrons in each
graphene sheet is given by eigenstates of the Hamiltonian which can be
written as a superposition of the sublattice basis. This implies that the
the detector,which acts on the sublattice basis will mix the eigenstates of
the Hamiltonian. For simplicity, we can write the initial state as $\rho
_{G}=\left\vert +,+\right\rangle \left\langle +,+\right\vert $ (see Fig. \ref%
{angles}) and thus%
\begin{equation}
\left\langle s^{\prime }\right\vert \sigma _{\lambda }\left\vert
s\right\rangle =\frac{1}{2}\left[ se^{i\theta }(1+\lambda )+s^{\prime
}e^{-i\theta ^{\prime }}(1-\lambda )\right]   \label{g1}
\end{equation}%
where the basis $\left\vert s\right\rangle $ is $\left\vert \pm
\right\rangle $. After a lenghty calculation $\left\langle s_{1}^{\prime
},s_{2}^{\prime }\right\vert \left( \sigma _{-\lambda }^{(i)}\sigma
_{-\lambda }^{(j)}\rho _{G}-2\sigma _{-\lambda }^{(i)}\rho _{G}\sigma
_{\lambda }^{(j)}+\rho _{G}\sigma _{\lambda }^{(i)}\sigma _{\lambda
}^{(j)}\right) \left\vert s_{1},s_{2}\right\rangle $ can be written as%
\begin{gather}
\left\langle s_{1}^{\prime },s_{2}^{\prime }\right\vert \left( \sigma
_{-\lambda }^{(i)}\sigma _{-\lambda }^{(j)}\rho _{G}-2\sigma _{-\lambda
}^{(i)}\rho _{G}\sigma _{\lambda }^{(j)}+\rho _{G}\sigma _{\lambda
}^{(i)}\sigma _{\lambda }^{(j)}\right) \left\vert s_{1},s_{2}\right\rangle =
\label{g2} \\
a_{11}\mathcal{F}_{11}+a_{22}\mathcal{F}_{22}+a_{12}\mathcal{F}_{12}  \notag
\end{gather}%
where $a_{11}=-\delta _{1s_{2}^{\prime }}\delta _{1s_{2}}e^{i(\theta
_{1}-\theta _{1}^{\prime })}(1+s_{1}s_{1}^{\prime })$, $a_{22}=-\delta
_{1s_{1}}\delta _{1s_{1}^{\prime }}e^{i(\theta _{1}-\theta _{1}^{\prime
})}(1+s_{2}s_{2}^{\prime })$ and%
\begin{eqnarray}
a_{12} &=&-e^{i(\theta _{1}-\theta _{2}^{\prime })}(\delta _{1s_{2}^{\prime
}}\delta _{1s_{1}}+\delta _{1s_{1}^{\prime }}\delta _{1s_{2}}s_{2^{\prime
}}s_{1})-e^{i(\theta _{2}-\theta _{1}^{\prime })}(\delta _{1s_{1}^{\prime
}}\delta _{1s_{2}}+\delta _{1s_{2}^{\prime }}\delta
_{1s_{1}}s_{2}s_{1}^{\prime })+  \label{g3} \\
&&e^{-i(\theta _{1}^{\prime }+\theta _{2}^{\prime })}(\delta
_{1s_{1}^{\prime }}\delta _{1s_{2}^{\prime }}+\delta _{1s_{1}}\delta
_{1s2}s_{1}^{\prime }s_{2}^{\prime })+e^{i(\theta _{1}+\theta _{2})}(\delta
_{1s_{1}}\delta _{1s2}+s_{1}s_{2}\delta _{1s_{1}^{\prime }}\delta
_{1s_{2}^{\prime }})  \notag
\end{eqnarray}%
where $\theta _{i}$($\theta _{i}^{\prime }$) is the initial (final) angle of
the wave vector $\mathbf{k_{i}}$ ($\mathbf{k_{i}^{\prime }}$), that appears
in the phase in Eq.(\ref{3}). The normalized reduced quantum operator reads
(see Eq.(8) of \cite{mar3})%
\begin{equation}
\rho =\left[ 
\begin{array}{cccc}
1+(ev_{F}t)^{2}\left[ A-D_{2}-D_{1}\right]  & -(ev_{F}t)^{2}B_{+} & 
-(ev_{F}t)^{2}B_{-} & (ev_{F}t)^{2}C \\ 
(ev_{F}t)^{2}B_{+} & (ev_{F}t)^{2}\left[ D_{2}-A\right]  & (ev_{F}t)^{2}E & 0
\\ 
(ev_{F}t)^{2}B_{-} & (ev_{F}t)^{2}E & (ev_{F}t)^{2}\left[ D_{1}-A\right]  & 0
\\ 
(ev_{F}t)^{2}C & 0 & 0 & (ev_{F}t)^{2}A%
\end{array}%
\right]   \label{g4}
\end{equation}%
where%
\begin{gather}
A=e^{-i(\theta _{1}^{\prime }+\theta _{2}^{\prime })}\left[ \mathcal{F}%
_{12}(-e^{i(\theta _{1}+\theta _{1}^{\prime })}-e^{i(\theta _{2}+\theta
_{2}^{\prime })}+e^{i(\theta _{1}^{\prime }+\theta _{2}^{\prime }+\theta
_{1}+\theta _{2})}+1)-e^{i(\theta _{2}^{\prime }+\theta _{1})}(\mathcal{F}%
_{11}+\mathcal{F}_{22})\right]   \label{g5} \\
B_{\pm }=\frac{1}{2}\mathcal{F}_{12}e^{-i(\theta _{1}^{\prime }+\theta
_{2}^{\prime })}(\mp 1+e^{i(\theta _{1}^{\prime }+\theta _{1})})(\pm
1+e^{i(\theta _{2}^{\prime }+\theta _{2})})  \notag \\
C=\frac{1}{2}\mathcal{F}_{12}(e^{-i(\theta _{1}^{\prime }+\theta
_{2}^{\prime })}+e^{i(\theta _{1}+\theta _{2})})  \notag \\
D_{i}=-\mathcal{F}_{ii}e^{-i(\theta _{1}^{\prime }-\theta _{1})}  \notag \\
E=-\frac{1}{2}\mathcal{F}_{12}(e^{i(\theta _{1}-\theta _{2}^{\prime
})}+e^{i(\theta _{2}-\theta _{1}^{\prime })})  \notag
\end{gather}%
From Eq.(\ref{g4}), the reduced operator is no longer an $X$ state due to
the matrix elements $B_{\pm }$, nevertheless, for specific choices of
initial and final angles of the wave vectors, different kinds of entangled
matrices can be obtained. For $B_{\pm }=0$ the angles must obey $\theta
_{1}^{\prime }+\theta _{1}=0$ and $\theta _{2}^{\prime }+\theta _{2}=0$ or $%
\theta _{1}^{\prime }+\theta _{1}=\pi $ and $\theta _{2}^{\prime }+\theta
_{2}=\pi $. In the first case, $A$ does not depend on $\mathcal{F}_{12}$ and
the matrix is identical to Eq.(\ref{r2}), but in the second case\ $%
e^{i(\theta _{1}+\theta _{1}^{\prime })}+e^{i(\theta _{2}+\theta
_{2}^{\prime })}-e^{i(\theta _{1}^{\prime }+\theta _{2}^{\prime }+\theta
_{1}+\theta _{2})}-1$ does not vanish and $\mathcal{F}_{12}$ appears in the
diagonal elements. In turn, when $\theta _{1}+\theta _{1}^{\prime }=\theta
_{2}+\theta _{2}^{\prime }-\pi $, $E=0$ and the density matrix can no longer
be related to all pure and mixed states by an entanglement-preserving
unitary transformation such that the transformed state has the same
entanglement as the input state, a property which is supported by strong
numerical evidence \cite{hede}. The correlated angles at which the electrons
in both layers scatter is related to the broken symmetry in double-layer
graphene shown in \cite{gor}. The matrix elements dependence of $\rho $ with
the initial and final angles implies that the nonlocal correlations are
sensitive to the relative orientation of the electrons.

An operational two-party entanglement-harvesting protocol to detect this
nonlocal correlation in double-layer graphene involves applying an external
voltage on both layers, which can vary the carrier concentration in the
material \cite{cra}. It is well known that graphene's density of states at
the neutral point vanishes, which implies that there are no states to occupy
and hence there are no carriers which could contribute to the electronic
transport. An usual procedure to change the charge concentration is to use
graphene as the second parallel plate of a capacitor, where the first plate
is SiO$_{2}$ and a back-gate voltage is applied perpendicular to the
graphene sheet which creates an electrostatic potential drop between the
sample and the gate electrode and shifts the Fermi level \cite{chakra}. The
distance between graphene layers should be an order of magnitude larger than
the capacitor in order to not change the boundary conditions for the
electromagnetic field used in the calculations \cite{bada}. By switching the
back-gate voltage\ on and off in one graphene layer in the interval $\left[
0,T_{1}\right] $ and performing the same procedure in the second layer in
the interval $\left[ T_{2},T_{2}^{\prime }\right] $ (where $T_{2}^{\prime
}-T_{2}=T_{1}$ and the initial time at which the second back-gate voltage is
turned on obeys $T_{2}<\frac{d_{1}-d_{2}}{c}$) and by measuring the current
in each graphene layer \cite{tsu}, it is possible to detect nonlocal
correlation even if both electrons do not exchange real photons.\footnote{%
The two voltages are switched on for the same amount of time but with a time
delay between them, which implies that the worldsheet of the second graphene
layer lies outside the light cone of the worldsheet of the first graphene
layer (see Fig. 1 of \cite{martsand}).} An improvement to the setup is to
introduce a dielectric in the whole cavity that changes the refractive index
and the velocity of light in order to decrease the time intervals at which
the back-gate voltages are switched on and off \cite{bada}.

Summing up, we have presented a new physical effect of vacuum fluctuations
which is associated with quantum nonlocality in double-layer graphene, which
allows to study relativistic quantum effects in the laboratory. It should be
stressed that this effect stands in contrast to other vacuum phenomena, such
as the Lamb shift or the Casimir effect \cite{farias}, which to some extent
can be emulated by classical stochastic local noise.

\section{Conclusions}

In this work we have performed a detailed study of the phenomenon of
entanglement harvesting from the vacuum state of the electromagnetic field
in double-layer graphene for different initial states for the electrons. By
considering that each graphene sheet interacts with the vacuum
electromagnetic field state and by partially tracing the degrees of freedom
of this field, the reduced quantum state of the electrons in different
layers gets entangled for times smaller than the time of flight of light
between the sheets. By using time-dependent perturbation theory up to second
order, the negativity measure of entanglement has been computed. We have
exhaustively analyzed the case in which both electrons are in one of the
pseudospin states, showing that for time scales smaller than the
light-crossing time between both layers, both electrons are correlated due
to the tails of the virtual photon propagator. In turn, we have shown that
when both electrons are in the conduction band, the reduced density matrix
reduces to an $X$ state for $\theta _{1}^{\prime }+\theta _{1}=0$ and $%
\theta _{2}^{\prime }+\theta _{2}=0$ or $\theta _{1}^{\prime }+\theta
_{1}=\pi $ and $\theta _{2}^{\prime }+\theta _{2}=\pi $ and for general
angles the bipartite quantum state becomes highly entangled with broken
electron-hole symmetry.

\section{Acknowledgment}

This paper was partially supported by grants from CONICET (Argentina
National Research Council) and Universidad Nacional del Sur (UNS) and by
ANPCyT through PICT 1770, and PIP-CONICET Nos. 114-200901-00272 and
114-200901-00068 research grants. J. S. A. is member of CONICET. J. S.A.
acknowledges financial support from CONICET (Argentina National Research
Council) to travel to the IMDEA Nanoscience Institute, Madrid, Spain during
2017-2018.

\appendix

\section{Appendix A}

In order to compute the photon propagator of Eq.(\ref{15})%
\begin{equation}
\left\langle \Omega \right\vert TA_{\lambda }^{(i)}(\mathbf{r_{1}}%
,t_{1})A_{\lambda ^{\prime }}^{(j)}(\mathbf{r_{2},}t_{2})\left\vert \Omega
\right\rangle =\delta _{\lambda \lambda ^{\prime }}\overset{\infty }{%
\underset{n=1}{\sum }}\gamma ^{2}\mathbf{\sin (}\frac{\pi nd_{i}}{L}\mathbf{%
)\sin (}\frac{\pi nd_{j}}{L}\mathbf{)}\int \frac{d^{2}\mathbf{q}}{(2\pi )^{2}%
}\frac{e^{i\mathbf{q(r_{1}-r_{2}})}e^{i\mathbf{\omega _{n,q}(t_{2}-t_{1})}}}{%
\omega _{n,\mathbf{q}}}  \label{a1}
\end{equation}%
we can apply the Schwinger time representation procedure by introducing a
new variable of integration $q_{0}$ 
\begin{equation}
\frac{e^{i\mathbf{\omega _{n,q}(t_{2}-t_{1})}}}{\omega _{n,\mathbf{q}}}%
=\int_{-\infty }^{\infty }\frac{dq_{0}}{2\pi i}\frac{2e^{iq_{0}\mathbf{%
(t_{2}-t_{1})}}}{q_{0}^{2}-q^{2}-(\frac{n\pi }{L})^{2}}  \label{a2}
\end{equation}%
The $q_{0}$ integration can be computed using the residue theorem and the
contour contains the $q_{0}$ real line and the semicircle of radius $R,$
where $R\rightarrow \infty $ and where the contour encloses the pole located
at $q_{0}=\omega _{n,\mathbf{q}}=\sqrt{q^{2}+(\frac{n\pi }{L})^{2}}$. Then,
we can apply the Wick rotation to the Euclidean space by defining $%
q_{0}=ip_{0}$ and $\mathbf{q=p}$, thus $d^{2}\mathbf{qdq_{0}=id^{3}p}$ and $%
q_{0}^{2}-\mathbf{q\cdot q}=-p_{0}^{2}-\mathbf{p^{2}}=-p^{2}$, and Eq.(\ref%
{a1}) becomes%
\begin{equation}
\int \frac{d^{2}\mathbf{q}}{(2\pi )^{2}}\int_{-\infty }^{\infty }\frac{dq_{0}%
}{2\pi i}\frac{2e^{i\mathbf{q(r_{1}-r_{2}})}e^{-iq_{0}\mathbf{(t_{1}-t_{2})}}%
}{q_{0}^{2}-q^{2}-(\frac{n\pi }{L})^{2}}\rightarrow -\int \frac{d^{3}\mathbf{%
p}}{(2\pi )^{3}}\frac{2e^{ip\cdot x}}{p^{2}+(\frac{n\pi }{L})^{2}}
\label{a3}
\end{equation}%
where $x=(\Delta t,-\Delta \mathbf{r})$. The last integral can be computed
by considering spherical coordinates $d^{3}\mathbf{p}=p^{2}dp\sin \theta
_{p}d\theta _{p}d\phi _{p}$ and by writing $p\cdot x=p\left\vert
x\right\vert \cos \theta _{p}$ where $\theta _{p}$ is the angle between the
momentum $\mathbf{p}$ and the vector $x$, $\left\vert x\right\vert =\sqrt{%
c^{2}\Delta t^{2}-\left\vert \Delta \mathbf{r}\right\vert ^{2}}$ and $\Delta 
\mathbf{r=r_{1}-r_{2}}$. Computing the integrals over $\theta _{p}$ and $%
\phi _{p}$, we obtain%
\begin{equation}
\Delta _{\lambda }^{(i,j)}(\Delta t,\Delta \mathbf{r})=-\frac{\delta
_{\lambda \lambda ^{\prime }}}{\pi ^{2}\left\vert x\right\vert }\overset{%
\infty }{\underset{n=1}{\sum }}\gamma ^{2}\mathbf{\sin (}\frac{\pi nd_{i}}{L}%
\mathbf{)\sin (}\frac{\pi nd_{j}}{L}\mathbf{)}\int_{0}^{\infty }\frac{%
pdp\sin (p\left\vert x\right\vert )}{p^{2}+(\frac{n\pi }{L})^{2}}  \label{a4}
\end{equation}%
Finally, the integral over $p$ reads%
\begin{equation}
\int_{0}^{\infty }\frac{pdp\sin (p\left\vert x\right\vert )}{p^{2}+(\frac{%
n\pi }{L})^{2}}=-i\theta (x^{2})\frac{1}{4\pi (x^{2}-i\epsilon )^{\frac{1}{2}%
}}e^{-i\frac{n\pi }{L}\sqrt{x^{2}-i\epsilon }}+\theta (-x^{2})\frac{1}{4\pi
(-x^{2}+i\epsilon )^{\frac{1}{2}}}e^{-\frac{n\pi }{L}\sqrt{-x^{2}+i\epsilon }%
}  \label{a5}
\end{equation}%
where we have used Eq. (27) of \cite{zhang} and finally the sum over $n$
reads%
\begin{gather}
\Delta _{\lambda }^{(i,j)}(\left\vert x\right\vert )=-\delta _{\lambda
\lambda ^{\prime }}\frac{\gamma ^{2}}{2\pi \left\vert x\right\vert }\overset{%
\infty }{\underset{n=1}{\sum }}\mathbf{\sin (}\frac{n\pi d_{i}}{L}\mathbf{%
)\sin (}\frac{n\pi d_{j}}{L}\mathbf{)}e^{-n\pi \frac{\left\vert x\right\vert 
}{L}}=  \label{a6} \\
\Delta _{\lambda }^{(i,j)}(\left\vert x\right\vert )=-\frac{\delta _{\lambda
\lambda ^{\prime }}}{16\pi \left\vert x\right\vert }\frac{\gamma ^{2}\sin (%
\frac{\pi d_{i}}{L})\sin (\frac{\pi d_{j}}{L})\sinh (\frac{\pi \left\vert
x\right\vert }{L})}{\sin (\frac{\pi (d_{i}-d_{j}-i\left\vert x\right\vert )}{%
2L})\sin (\frac{\pi (d_{i}+d_{j}-i\left\vert x\right\vert )}{2L})\sin (\frac{%
\pi (d_{i}-d_{j}+i\left\vert x\right\vert )}{2L})\sin (\frac{\pi
(d_{i}+d_{j}+i\left\vert x\right\vert )}{2L})}  \notag
\end{gather}%
which is the desired result for the photon propagator in the planar
microcavity.

\section{Appendix B}

In order to obtain Eq.(\ref{19}) we must compute the matrix elements of the
reduced density matrix $\rho (t)=Tr_{\phi }(U^{(2)}\rho _{0}+U^{(1)}\rho
U^{(1)\dag }+\rho _{0}U^{(2)\dag })$, that is $\left\langle \mathbf{%
k_{1}^{\prime },}s_{1}^{\prime }\mathbf{,k_{2}^{\prime },}s_{2}^{\prime
}\right\vert \rho (t)\left\vert \mathbf{k_{1},}s_{1}\mathbf{,k_{2},}%
s_{2}\right\rangle $, where $\mathbf{k_{i}^{\prime },}s_{i}^{\prime }$ are
the labels for the wave vector and $s=\pm 1$ the band index. It should be
noted that these matrix elements do not depend on the photon quantum states
due to the partial trace over these degrees of freedom. For simplicity we
will compute the matrix elements of the first term \ of $\rho (t)$, that is $%
\left\langle \mathbf{k_{1}^{\prime },}s_{1}^{\prime }\mathbf{,k_{2}^{\prime
},}s_{2}^{\prime }\right\vert Tr_{\phi }(U^{(2)}\rho _{0})\left\vert \mathbf{%
k_{1},}s_{1}\mathbf{,k_{2},}s_{2}\right\rangle $. We can write $Tr_{\phi
}(U^{(2)}\rho _{0})$ as

\begin{gather}
Tr_{\phi }(U^{(2)}\rho _{0})=-(ev_{F})^{2}\overset{}{\underset{%
i,j=1,2;\lambda ,\lambda ^{\prime }}{\sum }}\int_{0}^{t}%
\int_{0}^{t}dt_{1}dt_{2}\times   \label{ap1} \\
Tr_{\phi }\left[ e^{i(H_{0}+H_{F})t_{1}}\sigma _{-\lambda }^{(i)}A_{\lambda
}^{(i)}e^{-i(H_{0}+H_{F})t_{1}}e^{i(H_{0}+H_{F})t_{2}}\sigma _{-\lambda
^{\prime }}^{(j)}A_{\lambda ^{\prime }}^{(j)}e^{-i(H_{0}+H_{F})t_{2}}\rho
_{G}\right]   \notag
\end{gather}%
with $\rho _{0}=\left\vert \Omega \right\rangle \left\langle \Omega
\right\vert \rho _{G}$, where $\rho _{G}$ is the initial density operator of
the two-electron system. $\rho _{G}=\left\vert \mathbf{k_{1}^{(0)},}%
s_{1}^{(0)}\mathbf{,k_{2}^{(0)},}s_{2}^{(0)}\right\rangle \left\langle 
\mathbf{k_{1}^{(0)},}s_{1}^{(0)}\mathbf{,k_{2}^{(0)},}s_{2}^{(0)}\right\vert 
$, where $\mathbf{k_{i}^{(0)}}$ and $s_{i}^{(0)}$ are the initial wave
vectors and valence/conduction (or sublattice) indices. The last equation
can be written as%
\begin{equation}
Tr_{\phi }(U^{(2)}\rho _{0})=-(ev_{F})^{2}\overset{}{\underset{%
i,j=1,2;\lambda ,\lambda ^{\prime }}{\sum }}\int_{0}^{t}%
\int_{0}^{t}dt_{1}dt_{2}\left\langle \Omega \right\vert A_{\lambda
}^{(i)}(t_{1})A_{\lambda ^{\prime }}^{(j)}(t_{2})\left\vert \Omega
\right\rangle \sigma _{-\lambda }^{(i)}(t_{1})\sigma _{-\lambda ^{\prime
}}^{(j)}(t_{2})\rho _{G}  \label{ap2}
\end{equation}%
where $\sigma _{-\lambda }^{(i)}(t_{1})=e^{iH_{0}t_{1}}\sigma _{-\lambda
}^{(i)}e^{-iH_{0}t_{1}}$, $\sigma _{-\lambda ^{\prime
}}^{(j)}(t_{2})=e^{iH_{0}t_{2}}\sigma _{-\lambda ^{\prime
}}^{(j)}e^{-iH_{0}t_{2}}$, $A_{\lambda
}^{(i)}(t_{1})=e^{iH_{F}t_{1}}A_{\lambda }^{(i)}e^{-iH_{F}t_{1}}$ and $%
A_{\lambda ^{\prime }}^{(j)}(t_{2})=e^{iH_{F}t_{2}}A_{\lambda ^{\prime
}}^{(j)}e^{-iH_{F}t_{2}}$. We then apply $\left\langle \mathbf{k_{1}^{\prime
},}s_{1}^{\prime }\mathbf{,k_{2}^{\prime },}s_{2}^{\prime }\right\vert $ and 
$\left\vert \mathbf{k_{1},}s_{1}\mathbf{,k_{2},}s_{2}\right\rangle $ in the
coordinate representation%
\begin{gather}
\left\langle \mathbf{k_{1}^{\prime },}s_{1}^{\prime }\mathbf{,k_{2}^{\prime
},}s_{2}^{\prime }\right\vert Tr_{\phi }(U^{(2)}\rho _{0})\left\vert \mathbf{%
k_{1},}s_{1}\mathbf{,k_{2},}s_{2}\right\rangle =  \label{ap3} \\
-(ev_{F})^{2}\overset{}{\underset{i,j=1,2;\lambda ,\lambda ^{\prime }}{\sum }%
}\int d^{2}\mathbf{r_{1}}\int d^{2}\mathbf{r_{2}}\int_{0}^{t}%
\int_{0}^{t}dt_{1}dt_{2}e^{i\mathbf{k_{1}\cdot r_{1}}}e^{i\mathbf{k_{2}\cdot
r_{2}}}e^{-i\mathbf{k_{1}^{\prime }\cdot r_{1}}}e^{-i\mathbf{k_{2}^{\prime
}\cdot r_{2}}}\times   \notag \\
\left\langle \Omega \right\vert A_{\lambda }^{(i)}(\mathbf{r}_{1},d_{i}%
\mathbf{,}t_{1})A_{\lambda ^{\prime }}^{(j)}(\mathbf{r}_{2},d_{j}\mathbf{,}%
t_{2})\left\vert \Omega \right\rangle \left\langle s_{1}^{\prime
},s_{2}^{\prime }\right\vert \sigma _{-\lambda }^{(i)}(t_{1})\sigma
_{-\lambda ^{\prime }}^{(j)}(t_{2})\rho _{G}\left\vert
s_{1},s_{2}\right\rangle   \notag
\end{gather}%
In Appendix A it was shown that $\left\langle \Omega \right\vert TA_{\lambda
}^{(i)}(\mathbf{r_{1}},t_{1})A_{\lambda ^{\prime }}^{(j)}(\mathbf{r_{2},}%
t_{2})\left\vert \Omega \right\rangle =\Delta _{\lambda }^{(i,j)}(\left\vert
x\right\vert )=\delta _{\lambda \lambda ^{\prime }}F_{ij}(\left\vert
x\right\vert )$ where $\left\vert x\right\vert =\sqrt{\Delta
t^{2}-\left\vert \Delta \mathbf{r}\right\vert ^{2}}$. Then%
\begin{gather}
\left\langle \mathbf{k_{1}^{\prime },}s_{1}^{\prime }\mathbf{,k_{2}^{\prime
},}s_{2}^{\prime }\right\vert Tr_{\phi }(U^{(2)}\rho _{0})\left\vert \mathbf{%
k_{1},}s_{1}\mathbf{,k_{2},}s_{2}\right\rangle =  \label{ap4} \\
-(ev_{F})^{2}\overset{}{\underset{i,j=1,2;\lambda }{\sum }}%
\int_{0}^{t}\int_{0}^{t}\int d^{2}\mathbf{r_{1}}\int d^{2}\mathbf{r_{2}}%
dt_{1}dt_{2}e^{i(\mathbf{k_{1}-k_{1}^{\prime })\cdot r_{1}}}e^{i(\mathbf{%
k_{2}-k_{2}^{\prime })\cdot r_{2}}}F_{ij}(\left\vert x\right\vert
)\left\langle s_{1}^{\prime },s_{2}^{\prime }\right\vert \sigma _{-\lambda
}^{(i)}(t_{1})\sigma _{-\lambda }^{(j)}(t_{2})\rho _{G}\left\vert
s_{1},s_{2}\right\rangle   \notag
\end{gather}%
By performing the following change of variables $\mathbf{\Delta r=r_{1}-r_{2}%
}$, we have%
\begin{gather}
\left\langle \mathbf{k_{1}^{\prime },}s_{1}^{\prime }\mathbf{,k_{2}^{\prime
},}s_{2}^{\prime }\right\vert Tr_{\phi }(U^{(2)}\rho _{0})\left\vert \mathbf{%
k_{1},}s_{1}\mathbf{,k_{2},}s_{2}\right\rangle =  \label{ap5} \\
-(ev_{F})^{2}\overset{}{\underset{i,j=1,2;\lambda }{\sum }}%
\int_{0}^{t}\int_{0}^{t}dt_{1}dt_{2}\int d^{2}\mathbf{r_{1}}\int d^{2}%
\mathbf{r_{2}}e^{i(\mathbf{k_{1}-k_{1}^{\prime }+k_{2}-k_{2}^{\prime })\cdot
r_{1}}}e^{-i(\mathbf{k_{2}-k_{2}^{\prime })\cdot \Delta r}}\times   \notag \\
F_{ij}(\sqrt{\Delta t^{2}-\left\vert \Delta \mathbf{r}\right\vert ^{2}}%
)\left\langle s_{1}^{\prime },s_{2}^{\prime }\right\vert \sigma _{-\lambda
}^{(i)}(t_{1})\sigma _{-\lambda }^{(j)}(t_{2})\rho _{G}\left\vert
s_{1},s_{2}\right\rangle   \notag
\end{gather}%
Integrating over $\mathbf{r_{1}}$ we have%
\begin{gather}
\left\langle \mathbf{k_{1}^{\prime },}s_{1}^{\prime }\mathbf{,k_{2}^{\prime
},}s_{2}^{\prime }\right\vert Tr_{\phi }(U^{(2)}\rho _{0})\left\vert \mathbf{%
k_{1},}s_{1}\mathbf{,k_{2},}s_{2}\right\rangle =  \label{ap6} \\
-(ev_{F})^{2}\delta (\mathbf{k_{1}-k_{1}^{\prime }+k_{2}-k_{2}^{\prime }})%
\overset{}{\underset{i,j=1,2;\lambda }{\sum }}\int_{0}^{t}%
\int_{0}^{t}dt_{1}dt_{2}\left\langle s_{1}^{\prime },s_{2}^{\prime
}\right\vert \sigma _{-\lambda }^{(i)}(t_{1})\sigma _{-\lambda
}^{(j)}(t_{2})\rho _{G}\left\vert s_{1},s_{2}\right\rangle \mathcal{F}%
_{ij}\left( \mathbf{k_{2}-k_{2}^{\prime }}\right)   \notag
\end{gather}%
where $\mathcal{F}_{ij}\left( \mathbf{k_{2}-k_{2}^{\prime }}\right) $ is the
Fourier transform of $F_{ij}(\sqrt{\Delta t^{2}-\left\vert \Delta \mathbf{r}%
\right\vert ^{2}})$ 
\begin{equation}
\mathcal{F}_{ij}\left( \mathbf{k_{2}-k_{2}^{\prime },}\Delta t\right) =\int
d^{2}\mathbf{\Delta r}e^{-i(\mathbf{k_{2}-k_{2}^{\prime })\cdot \Delta r}%
}F_{ij}(\sqrt{\Delta t^{2}-\left\vert \Delta \mathbf{r}\right\vert ^{2}})
\label{ap7.0}
\end{equation}%
An identical procedure can be applied to $\left\langle \mathbf{k_{1}^{\prime
},}s_{1}^{\prime }\mathbf{,k_{2}^{\prime },}s_{2}^{\prime }\right\vert
Tr_{\phi }(U^{(1)}\rho _{0}U^{(1)\dag })\left\vert \mathbf{k_{1},}s_{1}%
\mathbf{,k_{2},}s_{2}\right\rangle $ and

$\left\langle \mathbf{k_{1}^{\prime },}s_{1}^{\prime }\mathbf{,k_{2}^{\prime
},}s_{2}^{\prime }\right\vert Tr_{\phi }(\rho _{0}U^{(2)\dag })\left\vert 
\mathbf{k_{1},}s_{1}\mathbf{,k_{2},}s_{2}\right\rangle $ and the results are
shown in Eq.(\ref{17}) and Eq.(\ref{18a}).

\bigskip

\end{document}